\documentclass[pre,twocolumn]{revtex4-1}
\usepackage{graphicx}
\usepackage{dcolumn}
\usepackage{bm}
\usepackage{amsmath}
\usepackage{amsfonts}

\begin{document}

\title{Elastic interactions between anisotropically contracting circular cells}

\author{Roman Golkov}
\email[]{romango1@tau.ac.il}
\author{Yair Shokef}
\email[]{shokef@tau.ac.il}
\homepage[]{https://shokef.tau.ac.il/}
\affiliation{School of Mechanical Engineering and The Sackler Center for Computational Molecular and Materials Science, Tel Aviv University, Tel Aviv 69978, Israel.}

\begin{abstract}
We study interactions between biological cells that apply anisotropic active mechanical forces on an elastic substrate. We model the cells as thin discs that along their perimeters apply radial, but angle-dependent forces on the substrate. We obtain analytical expressions for the elastic energy stored in the substrate as a function of the distance between the cells, the Fourier modes of applied forces and their phase angles. We show how the relative phases of the forces applied by the cells can switch the interaction between attractive and repulsive, and relate our results to those for linear force dipoles. For long enough distances, the interaction energy decays in magnitude as a power law of the cell-cell distance with an integer exponent that generally increases with the Fourier modes of the applied forces.
\end{abstract}

\maketitle

\section{Introduction}\label{introduction}

Mechanical forces influence the biological function at the cellular level. This is demonstrated most clearly in the experimental result that the differentiation of stem cells and their further fate depend on the mechanical properties of their environment~\cite{Engler2006, Fu2010, Kilian4872}. There is much current research also on the effects of mechanical forces on cell division~\cite{ayelet, Abuhattum2015}, embryonic development~\cite{Montell2003}, wound healing~\cite{Poujade2007}, cancer metastasis~\cite{Friedl2009, Gal2012}, cardiac beating~\cite{Nitsan2016}, and more. Specifically, living cells apply forces on their environment~\cite{Salbreux2012536}. The elastic properties of this environment largely affect the forces applied by the cells~\cite{Saez2005, Winer2009, Ghibaudo2008, Gilbert2010, Sazonova2011, Nisenholz2014, Shenoy2016}, the transmission of forces through the medium~\cite{lesman_arxiv}, their projected area~\cite{MASON20134635, King2010}, and interactions between distant cells~\cite{Reinhart-King2008, Shi2014}.

Cells connect to the extracellular matrix at focal adhesions, which are positioned on their surface. At these focused points cells apply on their mechanical environment  contractile forces, which are roughly directed toward the center of the cell~\cite{DemboWang1999, Balaban, Wang2002}. It was hence suggested that on a course-grained scale, the contractile activity of each cell may be modelled as a contractile force dipole~\cite{schwarz, SCHWARZ20021380, usam}. A \emph{linear force dipole} is a pair of opposing point forces of the same magnitude applied at some distance one from the other \cite{SafranSchwarz, Bischofs9274, PhysRevE.69.021911}. Each active cell generates a deformation field in the medium around it, which is in turn felt by distant cells, leading to a mechanical interaction~\cite{Bischofs2005, Bischofs2006}.

We previously studied spherically-contracting cells in a three-dimensional elastic medium. We focused first on the effects of the nonlinear material properties of the medium~\cite{Shokef2012, shokeferr}. We then introduced a mean-field approximation for interactions with neighboring cells~\cite{whycellscare}, and subsequently considered the full geometry of deformations induced by two spherical cells~\cite{1367-2630-19-6-063011}. For the latter case we identified an interaction mechanism that originates from shape regulation, and which does not exist for cells that do not regulate their mechanical activity due to the mechanical forces they sense~\cite{02, 03}, see also~\cite{Oakes2014}.

In this Paper we consider cells that adhere to the surface of an elastic substrate, and study how does their anisotropic contractility influence interactions between cells. Actual cells have irregular shapes, and we defer that to future work. Here, instead we limit ourselves to symmetric cells and focus on the effects of their anisotropic contractility on cell-cell interactions. Section~\ref{methods} describes our theoretical framework for describing the anisotropic displacement field generated by circular cells on an elastic substrate, and subsequently the interaction energy between them. In Section~\ref{results} we present our results for this interaction energy for different Fourier modes of anisotropic active forces. We find that for large enough distances, the interaction energy decays algebraically with distance with an integer exponent, which generally grows with the Fourier mode. In Section~\ref{discussion} we discuss our results, and specifically show how the relative phases of the Fourier modes in the two cells may cause the interaction energy to switch sign between attractive and repulsive. Section~\ref{conclusions} provides concluding remarks and outlook for future work.

\section{Methods}\label{methods}

\subsection{Theoretical framework}

\begin{figure}[t]
\includegraphics[width=\columnwidth]{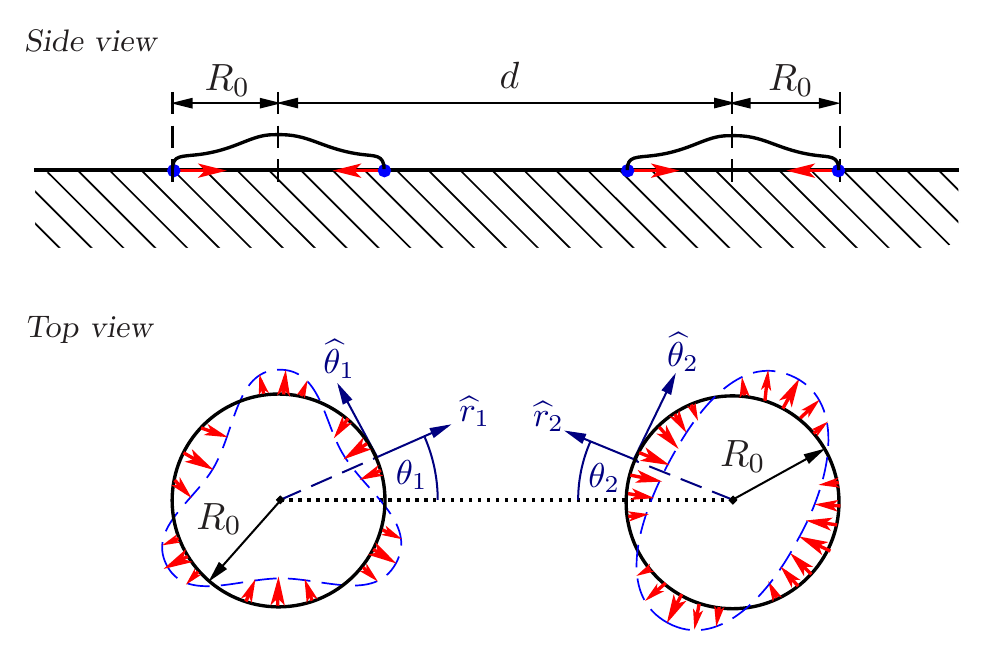}
\caption{Two active discs lying on a semi-infinite medium. We use a right-handed polar coordinate system for the left disc $1$ and a left-handed one for the right disc $2$. Angles $\gamma_1$ and $\gamma_2$ are the phases of the Fourier components of the active forces, $f_i(\theta)=C_{n,i}\cdot\cos\left[n_i\left(\theta_i-\gamma_i \right)\right]$. In the figure $n_1=3$, $n_2=2$.}
\label{two discs}
\end{figure}

We consider two \textit{active discs} each of radius $R_0$ lying on the surface of a semi-infinite elastic solid with linear Hookean behavior defined by bulk modulus $K$ and shear modulus $G$. We denote the distance between the centers of the discs by $d$, see Fig.~\ref{two discs}. Each disc is adhered to the surface of the underlying material along the disc's perimeter and applies there radial active forces on the substrate in an azimuthal distribution $f(\theta)$ which is not necessarily isotropic. While the nature of such an active disc is similar to that of a linear force dipole, there are several important differences between them. First, even though cell contractility typically does not produce a propulsive force, and cell propulsion requires additional processes, here we allow the net force applied by an active disc to be non-zero, namely the azimuthal distribution of active forces may include also a monopole component. Similarly, it may contain higher order multipole components, beyond the dipole moment that is contained in a linear force dipole.

As a result of the application of the active force, the underlying substrate deforms and forces are transduced from one disc to the other. We shall consider the interaction energy that we define here as the difference between the work performed by two interacting active discs in the system described above and the sum of self energies of two similar separate systems each of which includes only one such active disc. In the case we consider here of discs on a semi-infinite solid we expect the situation to be different than for three-dimensional spherical cells, in which in the absence of regulation, interaction energies vanish~\cite{02, 1367-2630-19-6-063011}. This is since for discs the displacement fields are not purely volume- or shape-changing anymore, as is the case for isotropically-contracting spheres. We restrict our present analysis to the case of radial forces. One can extend our work in a straightforward manner to consider active discs that apply also azimuthal forces, and analyze that situation using a similar method to what we employ here for the case of radial forces.

Since we assume linear elastic response of the substrate, by superposition we decompose the anisotropic force distribution that each disc generates into its Fourier components, and consider the interaction between two discs $i=1,2$ applying radial forces per unit length on their perimeters of the form $f_i(\theta_i)=\sum_n C_{n,i} \cos [n_{i} (\theta_i-\gamma_{n,i})]$, see Fig.~\ref{two discs}. Here $\theta_i$ are the polar angles for each disc, and $\gamma_{n,i}$ are the phases of all modes. Note that live cells apply only contractile (inward) forces. However, for the sake of our mathematical analysis, which treats each mode separately, we show in Fig.~\ref{two discs} a single Fourier component on each disc, and those single modes have both positive and negative forces since overall each mode has to be balanced. The total force $f(\theta)$ that a cell applies is the sum of multiple such modes and is always strictly positive (inward). This is typically obtained by having a positive $n=0$ mode.

We shall find the displacement fields created by each of the active discs and then sum them to get the total displacement, and subsequently from that we will obtain the interaction energy. We use a separate cylindrical coordinate system for each active disc with origin at its center. To simplify the calculations, we use the conventional right-handed polar coordinate system for active disc $1$, on the left, and a left-handed coordinate system for active disc $2$ on the right; namely the angle $\theta_2$ grows with rotation in a clockwise direction, see Fig.~\ref{two discs}.

\subsection{Displacement generated by active disc}

\begin{figure}[t]
\includegraphics[width=0.7\columnwidth]{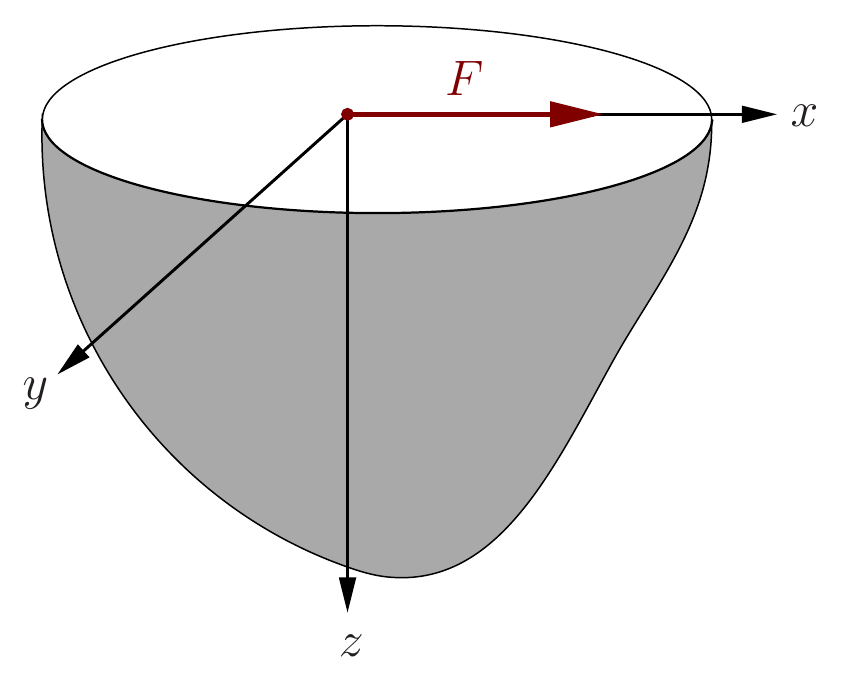}
\caption{Point force $F$ on the surface of the semi-infinite medium applied in the direction of the x-axis.} 
\label{Cherruti}
\end{figure}

To evaluate the interaction energy, we use the fact that it equals the additional work that is performed by the active discs in the presence of their neighbors. Since active discs apply forces only on the surface of the underlying solid and since work is given by an integral of total displacement times external force at the point of application of that force, we are only interested in the displacement on the surface of the semi-infinite solid. Furthermore, here all forces are coplanar with the free surface, thus we will only be interested in the displacements in the same plane and may neglect the displacements in the direction normal to the surface, see Fig.~\ref{Cherruti}.

The displacement field in the substrate must satisfy mechanical equilibrium, which we write in terms of the displacement field $\vec{u}$ as~\cite{01}:
\begin{equation}\label{Navier}
\frac{1}{1-2 \nu} \nabla \nabla \cdot \overrightarrow{u} + \nabla^2 \overrightarrow{u} = 0 
\end{equation}
where $\nu=\frac{3K-2G}{2(3K+G)}$ is the Poisson ratio of the medium. Due to the linearity of Eq.~\eqref{Navier} in the case of our Hookean medium, we may use the superposition principle, namely, we will decompose the angle-dependent forces created by the active discs to a system of point forces, solve the displacement field created by each of them and finally sum the displacements to find the resultant total displacement field. We will start with the Cerrutti Green's function~\cite{Johnson} for the displacement field due to a point force applied on the surface and in a direction tangent to the surface. For a force $F$ applied at the origin and directed along the x-axis (see Fig.~\ref{Cherruti}), the displacement is given by~\cite{Johnson, He2014116}:
\begin{align}
  u_x &= \frac{F}{4\pi G} \left[ \frac{1}{\rho} + \frac{x^2}{\rho^3} + (1-2\nu) \left\{ \frac{1}{\rho + z} - \frac{x^2}{\rho(\rho+z)^2} \right\} \right] , \label{point force displacement field 1}
\end{align}
\begin{align}
  u_y &= \frac{F}{4\pi G} \left[ \frac{x y}{\rho^3} - (1-2\nu) \frac{x y}{\rho (\rho+z)^2} \right]  , \label{point force displacement field 2}
\end{align}
\begin{align}
  u_z &= \frac{F}{4\pi G} \left[ \frac{x z}{\rho^3} - (1-2\nu) \frac{x}{\rho (\rho+z)} \right] , \label{point force displacement field 3}
\end{align}
where $\rho^2=x^2+y^2+z^2$. Since we are interested in plane displacements $\left( u_x, u_y \right)$, we will ignore Eq.~\eqref{point force   displacement field 3}. And since we are interested in the displacements on the surface, we will set $z=0$ in Eqs.~(\ref{point force displacement field 1}-\ref{point force displacement field 2}). For an active disc of radius $R_0$ applying a radial force per unit length $f(\theta)$ along its perimeter, Eqs.~(\ref{point   force displacement field 1}-\ref{point force   displacement field 2}) yield:
\begin{align}
  d u_x &= f(\theta) R_0 d\theta \frac{1}{2 \pi G \sqrt{x^2+y^2}} \left( 1 - \nu \frac{y^2}{x^2+y^2} \right) , \label{point force surface displacement field 1}\\
  d u_y &= f(\theta) R_0 d\theta \frac{\nu x y}{2 \pi G \left(x^2+y^2\right)^{3/2}} , \label{point force surface displacement field 2}
\end{align}
where $d u_x$ and $d u_y$ are the displacements created by the force applied along the arc $R_0 d\theta$. Here we assume that all the elastic response is due to the substrate. That is, we ignore the elastic resistance to deformation of each cell due to the forces applied by the other cell.

\begin{figure}[t]
\includegraphics[width=\columnwidth]{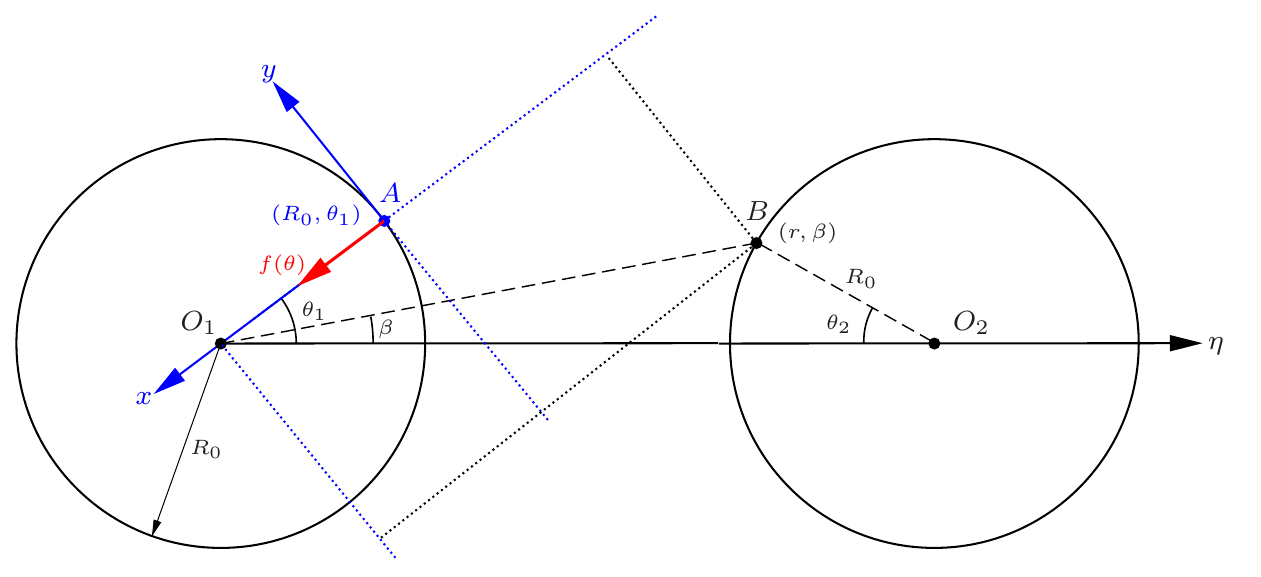}
\caption{Coordinate system used to evaluate the displacement at point B on the surface of active disc 2 as a result of the application of a radial point force F at point A on the boundary of active disc 1.} 
\label{singledisk}
\end{figure}

In order to evaluate the displacement field created by a single active disc we first rewrite the displacement field created by a point force in cylindrical coordinates with origin at the center of that active disc. Namely, as shown in Fig.~\ref{singledisk}, we assume a force per unit length $f_1(\theta_1)$, applied at point A on the perimeter of disc $1$ at orientation $\theta_1$ with respect to $\hat{\eta}$ -  the axis between the centers of the discs. We then evaluate the contribution of that force to the displacement at point B, that is located on the perimeter of disc $2$, namely on the surface of the medium, $z=0$, at a distance $r=\sqrt{d^2+R_0^2-2 d R_0 cos{\theta_2}}$ from the center of disc $1$, and with a polar angle $\beta=\arcsin{\left(R_0 sin{\theta_2}/r\right)}$ in disc $1$. Thus for the effect of point A on the displacement at point B, we should substitute in Eqs.~(\ref{point force   displacement field 1}-\ref{point force   displacement field 2}) above,
\begin{align}\label{cartesian to radial}
  x &= R_0 - r \cos{(\theta_1-\beta)} ,  \\
  y &= -r \sin{(\theta_1 - \beta)} . \label{cartesian to radial 2}
\end{align}
From Fig.~\ref{singledisk} and Eqs.~(\ref{cartesian to radial}-\ref{cartesian to radial 2}) it follows that:
\begin{align}
  d u_r &= -d u_x \cos (\theta_1-\beta) - d u_y \sin (\theta_1-\beta) \label{dur} , \\
  d u_{\theta} &= - d u_x \sin (\theta_1-\beta) + d u_y \cos (\theta_1-\beta) . \label{duth}
\end{align}
Here $du_r$ and $du_{\theta}$ are the radial and tangential displacements of point $B$ with respect to the origin $O_1$ due to the force $f R_0 d\theta$ applied by the azimuthal range $d\theta$ around point $A$, see Fig.~\ref{singledisk}. We substitute Eqs.~(\ref{point force surface displacement field 1}-\ref{point force surface displacement field 2}) in Eqs.~(\ref{dur}-\ref{duth}) and get:
\begin{widetext}
\begin{align}
 d \tilde{u}_r &= - \tilde{f}(\theta_1) \frac{ 2 \left(\tilde{r}^2+1 \right) \cos\left(\theta_1 -\beta\right)- \tilde{r} \left[2+\nu +\left(2-\nu \right) \cos \left(2 \left(\theta_1-\beta \right)\right)\right] }{4 \pi  \left[\tilde{r}^2-2 \tilde{r}\cos\left(\theta_1 -\beta\right)+1\right]^{3/2}} d\theta_1 , \label{dur dimensionless} \\
  d \tilde{u}_{\theta} &= -\tilde{f}(\theta_1) \frac{ \left[\tilde{r}^2 \left(1-\nu \right)- \tilde{r} \left(2-\nu \right) \cos \left(\theta_1 -\beta \right)+1\right] \sin \left(\theta_1 -\beta\right)}{2 \pi \left[\tilde{r}^2-2 \tilde{r}\cos\left(\theta_1 -\beta \right)+1\right]^{3/2}} d\theta_1 , \label{duth dimensionless}
\end{align}
\end{widetext}
where we have defined the dimensionless displacement $\tilde{u}_{r}=\frac{u_r}{R_0}$, $\tilde{u}_{\theta}=\frac{u_\theta}{R_0}$,  dimensionless force per unit length $\tilde{f_i}(\theta_i)=\frac{f_i(\theta_i)}{G R_0}$ and dimensionless position $\tilde{r}=\frac{r}{R_0}$. In order to evaluate the total displacement at point $B$ by superposition we integrate the contributions from all point forces $f(\theta_1)d\theta_1$ along the perimeter of active disc 1:
\begin{align}\label{u total}
  \vec{\tilde{u}} &= \int_{0}^{2 \pi} \vec{d \tilde{u}} ,
\end{align}
with vector notation: $\vec{\tilde{u}}=\left(
                            \begin{array}{c}
                              \tilde{u}_r \\
                              \tilde{u}_\theta \\
                            \end{array}
                          \right) , \vec{d \tilde{u}}=\left(
                                           \begin{array}{c}
                                             d \tilde{u}_r \\
                                             d \tilde{u}_\theta \\
                                           \end{array}
                                         \right)$.

\subsection{Interaction energy}

We compute the interaction energy from the additional work performed by each of the two active discs due to the presence of the neighboring active disc. To evaluate the interaction energy, we subtract from the total work the sum of the work that the same active discs would perform if there were no neighboring discs around them:
\begin{align}\label{interaction energy 2 discs}
  \Delta E &= E - \sum_{i=1}^2 E_0^{(i)} \\
  &= \frac{1}{2} \left[\sum_{i=1}^2 \int_{L_i} \left(\vec{u}_{i,tot} \cdot \vec{f}_i\right) dL_i \right. \nonumber \\
   &\left. \qquad\qquad\qquad\quad\quad-\sum_{i=1}^2 \int_{L_i} \left( \vec{u}_{0}^{(i)} \cdot \vec{f}_i \right) dL_i \right] \nonumber\\
   &=\frac{1}{2}\sum_{i=1}^2 \int_{L_i}\left( (\vec{u}_{i,tot}-\vec{u}_{0}^{(i)}) \cdot \vec{f}_i \right) dL_i , \nonumber
\end{align}
where $E$ is the total elastic energy stored in the medium for two interacting active discs and $E_0^{(i)}$ is the self energy of active disc $i$, i.e. the elastic energy of a system that consists only of this active disc. In addition, $\vec{u}_{i,tot}$ is the total displacement at the perimeter of active disc $i$, $\vec{f}_i$ is the force applied by it, $\vec{u}_{0}^{(i)}$ is the displacement that the force $\vec{f}_i$ applied by active disc $i$ would create in absence of the neighboring active disc, $L_i$ is the perimeter of active disc $i$ and $dL_i$ is the coordinate along it. From the last expression we see that the interaction energy is related to the product of the force applied by each disc multiplied by the displacement generated on the surface of that disc by the other disc. The coefficient $\frac{1}{2}$ in Eq.~\eqref{interaction energy 2 discs} may be explained in the following way~\cite{whycellscare}: each of the active discs eventually applies some force $\vec{f}d\theta$ at each point along its edge, and the total displacement at that point is eventually $\vec{u}$. We can think of an adiabatic process during which this force was built linearly in time over a duration $T$ such that at any time $0<t<T$ the force is given by: $\vec{g}(t)d\theta=\frac{t}{T} \vec{f}d\theta$. Then, by linearity of the medium, the displacement was built at the same rate. Namely, the displacement field at any time $t$ is given by $\vec{\omega}(t)=\frac{t}{T}\vec{u}$. Thus the work done in this process of building the force $\vec{f}$ and the displacement $\vec{u}$ is:
\begin{align}\label{Energyexplanation}
  dW&=\int_0^u \vec{g}(t)d\theta\cdot d\vec{\omega}(t) \\
  &\quad=\frac{1}{T^2} \int_{0}^{T} t dt \cdot \vec{f} R_0 d\theta\cdot \vec{u}=\frac{1}{2} \vec{f}R_0 d\theta\cdot\vec{u} . \nonumber
\end{align}

We now define the non-dimensional interaction energy as $\Delta\tilde{E}=\frac{\Delta E}{G R_0^3}$ and rewrite Eq.~\eqref{interaction energy 2 discs} as:
\begin{equation}\label{interaction energy general nond}
  \Delta \tilde{E} = \frac{1}{2} \left[\sum_{i=1}^2 \int_{2\pi} \left(\vec{\tilde{u}}_{i,tot}- \vec{\tilde{u}}_{0}^{(i)} \right) \cdot \vec{\tilde{f}}_i d\theta_i \right] .
\end{equation}
Note that $\Delta\tilde{E}$ in not normalized by the self energy $E_0$ but by the typical energy scale in the system, which we construct from the shear modulus of the substrate and the disc radius. According to Eq.~\eqref{interaction energy general nond} only the displacements along the perimeters of the active discs are relevant to the computation of the interaction energy. The total displacement around active disc $i$ is:
\begin{equation}\label{total displacement 2 discs}
  \vec{\tilde{u}}_{i,tot}=\vec{\tilde{u}}_{i}+ \vec{\tilde{u}}_{ji}=\vec{\tilde{u}}_{0}^{(i)}+\vec{\tilde{u}}_{ji} .
\end{equation}
Here $\vec{\tilde{u}}_{i}$ is the self-displacement created by disc $i$ at its perimeter while displacement $\vec{\tilde{u}}_{ji}$ is created at the same region by neighboring disc $j$. We consider the case that the forces $\vec{\tilde{f}}_i$ applied by the active discs do not depend on the presence of neighboring discs. Thus the self-displacement created  by each active disc does not depend on the presence of neighboring discs, i.e. $\vec{\tilde{u}}_i=\vec{\tilde{u}}_{0}^{(i)}$. Substitution of Eq.~\eqref{total displacement 2 discs} in Eq.~\eqref{interaction energy general nond} leads to~\cite{ESHELBY195679}:
\begin{align}\label{interaction energy simplified 2 discs}
  \Delta \tilde{E} &= \frac{1}{2} \left[\sum_{i=1}^2 \int_{2 \pi} \left(\vec{\tilde{u}}_{i,tot} -  \vec{\tilde{u}}_{0}^{(i)} \right) \vec{\tilde{f}}_i  d\theta_i\right] \\
  &=\frac{1}{2} \sum_{i=1}^2 \int_{2\pi} \vec{\tilde{u}}_{ji} \cdot \vec{\tilde{f}}_i d\theta_i \nonumber \\
  &=\frac{1}{2}\left[\int_{2\pi} \vec{\tilde{u}}_{21} \cdot \vec{\tilde{f}}_1d\theta_1 +\int_{2\pi} \vec{\tilde{u}}_{12} \cdot \vec{\tilde{f}}_2 d\theta_2\right] . \nonumber
\end{align}

In the general case of a system of $N$ active discs  Eq.~\eqref{interaction energy simplified 2 discs} reads:
\begin{equation}\label{interaction energy simplified}
  \Delta \tilde{E} =\frac{1}{2} \sum_i \int_{2\pi} \sum_{j\ne i} \vec{\tilde{u}}_{ji} \cdot \vec{\tilde{f}}_i d\theta_i=\sum_{i\ne j} \frac{1}{2} \int_{2\pi} \vec{\tilde{u}}_{ji} \cdot \vec{\tilde{f}}_i d\theta_i .
\end{equation}
So, an alternative way of writing the interaction energy is:
\begin{equation}\label{alternative interaction energy}
  \Delta \tilde{E} = \sum_{i\ne j}\Delta \tilde{E}_{ij} ,
\end{equation}
where:
\begin{equation}\label{deltaEij}
  \Delta \tilde{E}_{ij} = \frac{1}{2} \int_{2\pi} \vec{\tilde{u}}_{ji} \cdot \vec{\tilde{f}}_i d\theta_i
\end{equation}
is the interaction energy of active disc $i$ with active disc $j$, or formulating it another way, it is the amount of additional work that active disc $i$ performs in the presence of active disc $j$. Since $\vec{f}$ is in the radial direction, the product $\vec{\tilde{u}}_{ji} \cdot \vec{\tilde{f}}_i$ in Eq.~\eqref{alternative interaction energy} becomes $u_r f$ so we need only the radial part of the total displacement. However note that we need the component of the displacement that is directed toward the center of disc $i$, which when analyzed in the coordinate system of disc $j$ has both radial and azimuthal components there. It is important to emphasize that $\tilde{E}_{ij}$ and $\tilde{E}_{ji}$ have different physical meanings and are generally not necessarily equal.

\begin{figure}[b]
\includegraphics[width=\columnwidth]{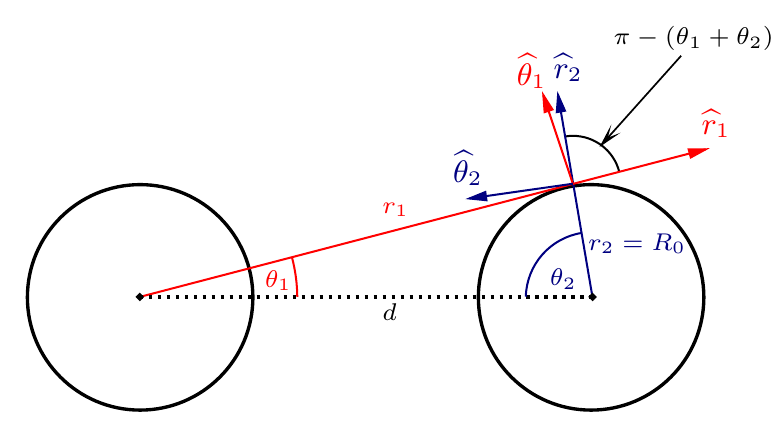}
\caption{Transformation of the coordinate system $(r_1,\theta_1)$ to the coordinate system $(r_2,\theta_2)$.} 
\label{transformation}
\end{figure}

The interaction energy between two active discs is $\Delta \tilde{E}=\Delta \tilde{E}_{12} + \Delta \tilde{E}_{21}$. We use Eq.~\eqref{deltaEij} to compute the interaction energies $\Delta \tilde{E}_{12}$ and $\Delta \tilde{E}_{21}$. To evaluate the integral in Eq.~\eqref{deltaEij} we first need to transform the displacement field generated by disc $j$ from its coordinate system to the coordinate system of disc $i$. The transformation is done by multiplying the vector $\vec{d \tilde{u}_j}$, Eqs.~(\ref{dur dimensionless}-\ref{duth dimensionless}) by the appropriate rotation matrix $B^{(ji)}$. Here $j$ and $i$ in the superscript denote the original and the target coordinate systems, respectively. For the system depicted in Fig.~\ref{transformation} the rotation matrix $B^{(ji)}$ is given by:
\begin{equation}\label{rotation matrix}
  B^{(12)}=B^{(21)}=  \left(
                  \begin{array}{cc}
                    -cos(\theta_1 + \theta_2) & sin(\theta_1 + \theta_2) \\
                     sin(\theta_1 + \theta_2)  & cos(\theta_1 + \theta_2)
                  \end{array}
                \right) ,
\end{equation}
where the first equality follows from our choice of left- and right-handed coordinate systems, see Fig.~\ref{transformation}. The resultant displacement field is:
\begin{equation}\label{rotated displacements}
  \vec{\tilde{u}}_{12}=\left(
                               \begin{array}{c}
                                  -cos(\theta_1 + \theta_2) \tilde{u}_r +  sin(\theta_1 + \theta_2) \tilde{u}_\theta \\
                                 sin(\theta_1 + \theta_2)  \tilde{u}_r  +  cos(\theta_1 + \theta_2) \tilde{u}_\theta
                               \end{array}
                             \right) .
\end{equation}
Since the forces applied by each of the active discs are directed in the radial direction, it follows from Eq.~\eqref{deltaEij} that we only need the first, radial component in Eq.~\eqref{rotated displacements}. The interaction energy $\Delta \tilde{E}_{21}$ then becomes:
\begin{align}\label{interaction energy 12}
  \Delta \tilde{E}_{21} &= \frac{1}{2} \int_{-\pi}^{\pi} \left( \vec{\tilde{u}}_{12} \cdot \vec{\tilde{f}}_2 \right) d\theta_2 =\\
  &= \frac{1}{2}\int_{-\pi}^{\pi} \left[ -cos(\theta_1 + \theta_2) \tilde{u}_r \right. \nonumber\\
  &\qquad\qquad\qquad\left. + sin(\theta_1 + \theta_2) \tilde{u}_\theta \right] \tilde{f}_2 (\theta_2) d \theta_2. \nonumber
\end{align}
As mentioned above, the interaction energy $\Delta \tilde{E}_{21}$ depends only on the radial part of the displacement $\vec{\tilde{u}}_{12}$, see Eqs.~(\ref{rotation matrix}-\ref{rotated displacements}). The angle $\theta_1$ in this expression originally belongs to the rotation matrix $B^{ji}$ (see Eq.~(\ref{rotation matrix}-\ref{rotated displacements})) and will be replaced by the following expressions in terms of $\theta_2$:
\begin{align}\label{theta substitution}
  \tilde{r} &= \sqrt{\tilde{d}^2+1-2\tilde{d}\cos{\theta_2}} , \\
  \sin{(\theta_1)} &= \frac{\sin{\theta_2}}{\tilde{r}} , \\
  \cos{(\theta_1)} &=  \frac{\tilde{d}-\cos{\theta_2}}{\tilde{r}} , \label{theta substitution end}
\end{align}
where we have introduced the dimensionless distance between the cells $\tilde{d}=d/R_0$. In order to get $\Delta \tilde{E}_{21}$ using Eq.~\eqref{interaction energy 12} we need to compute $\tilde{u}_r$ and $\tilde{u}_\theta$, which are the radial and tangential (with respect to $O_1$) components of the displacement field created by active disc $1$ along the perimeter of active disc $2$. We evaluate them by using trigonometric relations for $\sin{(\theta-\beta)}$ and $\cos{(\theta-\beta)}$ in Eqs.~(\ref{dur dimensionless}-\ref{u total}) and then substituting the following relations:
\begin{align}\label{alpha beta substitution}
\tilde{r} &= \sqrt{\tilde{d}^2+1-2\tilde{d}\cos{\theta_2}} , \\
\sin{(\beta)} &= \frac{\sin{\theta_2}}{\tilde{r}} , \\
\cos{(\beta)} &=  \frac{\tilde{d}-\cos{\theta_2}}{\tilde{r}}, \label{alpha beta substitution end}
\end{align}
which follow from the cosine theorem for the geometry of the system of two active discs, as shown in Fig.~\ref{singledisk}. See also Eqs.~(\ref{theta substitution}-\ref{theta   substitution end}). We integrate over $\theta_1$ in accordance with Eqs.~(\ref{dur  dimensionless}-\ref{u total}) and substitute the results that do not depend on $\theta_1$  anymore in Eq.~\eqref{interaction energy 12} that includes instances of $\theta_1$ that come from the rotation matrix $B^{(21)}$, see Eqs.~(\ref{rotation matrix}-\ref{interaction energy 12}). In order to complete the coordinate transformation from $(r_1, \theta_1)$ to $(r_2,\theta_2)$ we allow the same procedure that we employed for $\beta$ to these remaining instances of $\theta_1$. Namely, we substitute in Eq.~\eqref{interaction energy 12}  $\sin{\theta_2}/\tilde{r}$ and $(\tilde{d}-\cos{\theta_2})/\tilde{r}$ for $\sin(\theta_1)$ and $\cos(\theta_1)$, respectively, in accordance with Fig.~\ref{transformation} and with Eqs.~(\ref{theta substitution}-\ref{theta   substitution end}). After substitution and integration over $\theta_2$ we evaluate the expression for $\Delta \tilde{E}_{21}$. In the same manner we evaluate $\Delta \tilde{E}_{12}$ and then in accordance with Eq.~\eqref{alternative interaction energy} we sum $\Delta \tilde{E}_{21}$ and  $\Delta \tilde{E}_{21}$ to get the expression for the total interaction energy in this system.

\section{Results}\label{results}

\subsection{Interaction between single Fourier modes}

If we take $f_1(\theta_1)=A_1 \cos [n (\theta_1-\gamma_1)]$ and $f_2(\theta_2)=A_2 \cos [m (\theta_2-\gamma_2)]$, then from Eqs.~(\ref{dur dimensionless}-\ref{duth   dimensionless}) we get:
\begin{widetext}
\begin{align}
 &\left(d \tilde{u}_r\right)_n = - a_1 \cos [n (\theta_1-\gamma_1)] \frac{ \left\{2 \left(\tilde{r}^2+1 \right) \cos\left(\theta_1 -\beta\right)- \tilde{r} \left[2+\nu +\left(2-\nu \right) \cos \left(2 \left(\theta_1-\beta \right)\right)\right] \right\}}{4 \pi  \left[\tilde{r}^2-2 \tilde{r}\cos\left(\theta_1 -\beta\right)+1\right]^{3/2}} d\theta_1 , \label{dur dimensionless n} \\
 &\left(d \tilde{u}_{\theta}\right)_n = - a_1 \cos [n (\theta_1-\gamma_1)] \frac{ \left[\tilde{r}^2 \left(1-\nu \right)- \tilde{r} \left(2-\nu \right) \cos \left(\theta_1 -\beta \right)+1\right] \sin \left(\theta_1 -\beta\right)}{2 \pi \left[\tilde{r}^2-2 \tilde{r}\cos\left(\theta_1 -\beta \right)+1\right]^{3/2}} d\theta_1 . \label{duth dimensionless n}
\end{align}
\end{widetext}
Here $a_i = \frac{A_i}{G R_0}$ is the dimensionless magnitude of the force, and the subscript $n$ indicates the harmonic mode of the force. After substitution of Eqs.~(\ref{alpha beta substitution}-\ref{alpha beta substitution end}) in Eqs.~(\ref{dur  dimensionless n}-\ref{duth  dimensionless n}) we get:
\begin{align}
\left(d \tilde{u}_r\right)_n &= a_1 \cos [n (\theta_1-\gamma_1)] \frac{\varphi}{2 \kappa} d\theta_1 , \\
\left(d \tilde{u}_{\theta}\right)_n &= a_1 \cos [n (\theta_1-\gamma_1)]  \frac{\psi}{\kappa} d\theta_1 ,
\end{align}
with
\begin{align}
\varphi &\equiv (2+\nu)\left(1+\tilde{d}^2\right) -2 \tilde{d} \left(\tilde{d}^2+3\right) \cos \left(\theta_1\right)\\
  &+\tilde{d}^2 \left(2- \nu \right) \cos \left(2 \theta _1\right) +2 \tilde{d}^2 \cos \left(\theta _1-\theta _2\right) \nonumber \\
  &-2 \tilde{d}(2+ \nu ) \cos
   \left(\theta _2\right)+4\left(1+ \tilde{d}^2\right) \cos \left(\theta _1+\theta
   _2\right) \nonumber\\
   &-2\tilde{d} (2-\nu ) \cos \left(2 \theta _1+\theta _2\right) \nonumber \\
   &-2 \tilde{d}
   \cos \left(\theta _1+2 \theta _2\right)+(2-\nu ) \cos \left(2 \theta_1+2 \theta _2\right) , \nonumber
\end{align}
\begin{align}
   \psi &\equiv \left[\tilde{d} \sin \left(\theta _1\right)-\sin \left(\theta _1+\theta
   _2\right)\right] \left[-2-\tilde{d}^2+\nu \right.  \\
   &+\tilde{d}^2 \nu +\tilde{d} (2-\nu ) \cos \left(\theta_1\right)+2 \tilde{d} (1-\nu ) \cos \left(\theta _2\right) \nonumber \\
   &\left. -(2-\nu ) \cos\left(\theta _1+\theta _2\right)\right] \nonumber ,
\end{align}
and
\begin{align}
   \kappa &\equiv 2 \pi  \sqrt{1+\tilde{d}^2-2 \tilde{d} \cos \left(\theta _2\right)} \left[2+\tilde{d}^2-2 \tilde{d} \cos
   \left(\theta _1\right) \right.\\
   & \left. -2 \tilde{d} \cos \left(\theta _2\right)+2 \cos \left(\theta _1+\theta _2\right)\right]^{3/2} . \nonumber
\end{align}
On substitution of these expressions in Eq.~\eqref{u total} and integration with respect to $\theta_1$ we find the displacement field $\vec{\tilde{u}}^{(1)}_n$, where the superscript $1$ corresponds to the active disc that created it. In order to simplify the resultant expressions we solve for the case $\tilde{d}\gg 1$ of active discs separated by a distance which is substantially larger than their radius. Further substitution in Eq.~\eqref{interaction energy 12} gives:
\begin{align}\label{interaction energy 12nm}
  \Delta \tilde{E}_{21,nm}
  &= \frac{1}{2}\int_{-\pi}^{\pi} \left[ -\cos(\theta_1 + \theta_2) \left(\tilde{u}_r\right)_n^{(1)} \right.  \\
   & \qquad \left. + \sin(\theta_1 + \theta_2) \left(\tilde{u}_\theta\right)_n^{(1)} \right] \cos(m \theta_2)d \theta_2. \nonumber
\end{align}
Here $\Delta \tilde{E}_{21,nm}$ is the amount of additional work that active disc $2$ performs due to the presence of active disc $1$ if they apply radial forces $\tilde{f}_2(\theta_2)=a_2 \cos [m (\theta_2 - \gamma_2)]$ and $\tilde{f}_1(\theta_1)= a_1 \cos [n (\theta_1 - \gamma_1)]$, respectively.

\subsection{Isotropic forces}

As a demonstration of this procedure we take a simple case, in which the forces applied by the active discs are isotropic, i.e. $n=m=0$ and thus $\tilde{f}_1(\theta_1)=\tilde{f}_2(\theta_2)\equiv 1$. In this case Eqs.~(\ref{dur  dimensionless}-\ref{duth   dimensionless}) become:
\begin{widetext}
\begin{align}
 d \tilde{u}_r &=- \frac{ \left\{2 \left(\tilde{r}^2+1 \right) \cos\left(\theta -\beta\right)- \tilde{r} \left[2+\nu +\left(2-\nu \right) \cos \left(2 \left(\theta-\beta \right)\right)\right] \right\}}{4 \pi  \left[\tilde{r}^2-2 \tilde{r}\cos\left(\theta -\beta\right)+1\right]^{3/2}} d\theta , \label{dur0mode}  \\
  d \tilde{u}_{\theta} &= -\frac{ \left[\tilde{r}^2 \left(1-\nu \right)- \tilde{r} \left(2-\nu \right) \cos \left(\theta -\beta \right)+1\right] \sin \left(\theta -\beta\right)}{2 \pi \left[\tilde{r}^2-2 \tilde{r}\cos\left(\theta -\beta \right)+1\right]^{3/2}} d\theta . \label{duth0mode}
\end{align}
\end{widetext}
Here the symmetry of the applied forces results in the cancellation of $u_{\theta}$, and moreover $u_r$ does not depend on $\beta$. Thus we cancel $\beta$ in Eq.~\eqref{dur0mode}, substitute it in Eq.~\eqref{u total} and get
\begin{align}\label{durtotmod0}
    \tilde{u}_r&=-\frac{(1-\nu)}{\pi  \tilde{r} (1+\tilde{r})}\left[\left(1+\tilde{r}^2\right) K\left(k\right) -(1+\tilde{r})^2 E\left(k\right)\right] ,
\end{align}
which is consistent with \cite{Johnson}. Here $K(k)$ and $E(k)$ are the complete elliptic integrals of the first and the second kind, respectively~\cite{ElFunc}:
\begin{align}\label{elliptic integrals}
  K(x) &= \int_{0}^{\frac{1}{2}\pi} \sqrt{1-x^2 \sin^2 \theta} d\theta, \\
  E(x) &= \int_{0}^{\frac{1}{2}\pi}\frac{d\theta}{ \sqrt{1-x^2 \sin^2 \theta}},
\end{align}
and $k=\frac{2\sqrt{\tilde{r}}}{1+\tilde{r}}$. Since both active discs apply the same forces that do not depend on the phases $\gamma_1$ and $\gamma_2$, the total interaction energy equals $\Delta E = 2 \Delta E_{21}$ where $\Delta E_{21}$ is given by Eq.~\eqref{interaction energy 12}. Since Eqs.~(\ref{alpha beta substitution}-\ref{alpha beta substitution end}) have to be substituted in Eq.~\eqref{durtotmod0}, the resultant expression becomes too cumbersome to be integrated analytically in accordance with Eq.~\eqref{u total}. For higher modes $(n,m>0)$ the equations become even more complicated and analytic solutions for the integrals in Eqs.~(\ref{u total}) and (\ref{interaction energy 12}) are not known. To overcome this difficulty,  we approximate for large separations $\tilde{d}\gg 1$ the expressions for $\vec{d \tilde{u}}$ and $\vec{\tilde{u}}^{(12)}$ using a Taylor series expansion to leading order in $1/\tilde{d}$.

To demonstrate this in our $n=m=0$ example, we set $a_1=a_2=1$, substitute Eqs.~(\ref{alpha beta substitution}-\ref{alpha beta substitution end}) in Eqs.~(\ref{dur0mode}-\ref{duth0mode}), approximate the result to leading order in $1/\tilde{d}$ and obtain:
\begin{align}
 &d \tilde{u}_{r} = -\frac{1}{16 \tilde{d}^3 \pi} \left[ 4 \tilde{d}(1-\nu)  \right.  \label{dur0approx}\\
   &\qquad +\left(8 \tilde{d}^2+5-6 \nu \right) \cos \left(\theta_1\right)+4 \tilde{d}(1+\nu ) \cos \left(2 \theta _1\right)\nonumber\\
   &\qquad +3(1+2 \nu ) \cos \left(3 \theta _1\right)+8 \cos \left(\theta _1-2 \theta _2\right) \nonumber \\
   &\qquad +8 \tilde{d}\cos \left(\theta _1-\theta _2\right)+8(1+\nu ) \cos \left(2 \theta
   _1-\theta _2\right) \nonumber\\
    &\qquad\left.+8(1- \nu )\vphantom{\tilde{d}} \cos \left(\theta _2\right) \right] d\theta_1 , \nonumber
\end{align}
\begin{align}
  &d \tilde{u}_{\theta} = -\frac{1}{16 \tilde{d}^3 \pi}  \left[ \left(8 \tilde{d}^2(1-\nu)-1+3 \nu \right) \sin \left(\theta _1\right) \right. \label{duth0approx} \\
  & \qquad +4\tilde{d}(1-2 \nu ) \sin \left(2 \theta _1\right)+3(1-3 \nu ) \sin \left(3 \theta_1\right) \nonumber\\
   &+8(1- \nu ) \sin \left(\theta _1-2 \theta _2\right)-8 \tilde{d}(1- \nu ) \sin \left(\theta _1-\theta _2\right) \nonumber \\
   & \left. +8(1-2 \nu ) \sin \left(2
   \theta _1-\theta _2\right) \vphantom{\tilde{d}} \right] d\theta_1. \nonumber
\end{align}
Note that in spite of the fact that $\theta_2$ is present in Eqs.~(\ref{dur0approx}-\ref{duth0approx}), $d \tilde{u}_{r}$ and $d \tilde{u}_{\theta}$ are the radial and azimuthal displacements in the coordinate system of active disc $1$. In order to rewrite them in the coordinate system of active disc $2$ one has to multiply them by the rotation matrix $B^{(12)}$, see Eq.~\eqref{rotation matrix}.
We integrate Eqs.~(\ref{dur0approx}-\ref{duth0approx}) according to Eq.~\eqref{u total} and get:
\begin{align}
  \tilde{u}_{r} &= - \frac{\left(1-\nu \right) \left(\tilde{d}+2 \cos \left(\theta_2\right)\right)}{2 \tilde{d}^3} , \label{ur0approx} \\
  \tilde{u}_{\theta} &= 0. \label{uth0approx}
\end{align}
Substitution of Eq.~\eqref{ur0approx}-\eqref{uth0approx} in Eq.~\eqref{rotated displacements} gives the expressions for the displacement field created by active disc $1$ rewritten in the coordinate system of active disc $2$. In order to evaluate the interaction energy we only need the radial component of that field:
\begin{equation}\label{ur21}
  \left(\tilde{u}_{12}\right)_r=\frac{\left(1-\nu\right) \left(\tilde{d}+2 \cos \left(\theta
   _2\right)\right) \left(\tilde{d} \cos \left(\theta
   _2\right)-1\right)}{2 \tilde{d}^3 \sqrt{\tilde{d}^2-2 \tilde{d} \cos \left(\theta
   _2\right)+1}} .
\end{equation}
The interaction energy in accordance with Eq.~\eqref{interaction energy 12} is:
\begin{equation}\label{interaction energy 0 mode}
  \Delta \tilde{E}_{21}= - \frac{\pi  \left(1-\nu\right)}{2 \tilde{d}^3},
\end{equation}
and due to the symmetry of the system $\Delta \tilde{E}= 2\Delta \tilde{E}_{21}$, and in dimensional form we have:
\begin{equation}
\Delta E = - \pi (1-\nu) G R_0^3 \cdot \frac{A_1}{GR_0} \cdot \frac{A_2}{GR_0} \cdot \left(\frac{R_0}{d}\right)^3 .
\end{equation}
This interaction energy clearly scales with the typical energy scale $G R_0^3$ in the system. It depends on the product of the dimensionless force magnitudes $a_i = \frac{A_i}{G R_0}$, and decays algebraically with the dimensionless distance $\tilde{d} = \frac{d}{R_0}$ between discs. We get $\Delta E<0$, which means that active discs will be attracted to each other. We will later show that for higher modes $\Delta E$ may be positive or negative depending on the phases $\gamma_1$, $\gamma_2$. Since the forces applied by the active discs in this case $(n=m=0)$ are isotropic, the resultant interaction energy does not include $\gamma_1$ and $\gamma_2$ while for any other modes of the forces, the interaction energy will depend on them. 

\subsection{Higher modes}

We find that the dimensionless interaction energies $\Delta \tilde{E}_{nm}$ between any modes $n$ and $m$ of the dimensionless forces $\tilde{f}_1(\theta_1) = a_1 \cos{[n(\theta_1-\gamma_1)]}$ and $\tilde{f}_2(\theta_2) = a_2 \cos{[m(\theta_2-\gamma_2)]}$ have the general form
\begin{align}\label{Enm}
&\Delta \tilde{E}_{nm} = -\frac{\pi a_1 a_2}{\tilde{d}^{|n-1|+|m-1|+1}}\\
&\cdot \left[(2-\nu)F_{nm}\cos(n\gamma_1 +m\gamma_2) + \nu H_{nm}\cos(n\gamma_1 -m\gamma_2)\right] , \nonumber
\end{align}
where the coefficients $F_{nm}$, $H_{nm}$ are given in Tables~\ref{G1G2}-\ref{CnCm} for the first 10 modes. In dimensional form this is given by
\begin{align}\label{Enm_dim}
&\Delta E_{nm} = - \pi G R_0^3 \cdot \frac{A_1}{GR_0} \cdot \frac{A_2}{GR_0} \cdot \left(\frac{R_0}{d}\right)^{|n-1|+|m-1|+1}\\
&\cdot \left[(2-\nu)F_{nm}\cos(n\gamma_1 +m\gamma_2) + \nu H_{nm}\cos(n\gamma_1 -m\gamma_2)\right] , \nonumber
\end{align}
where $f_i(\theta_i)$ and thus $A_i$ have dimension of force per unit length.

\begin{widetext}
\center
\begin{table}[!htb]
  \centering
\begin{tabular}{|c|c|c|c|c|c|c|c|c|c|c|}
 \hline
 $m \backslash n$
  & $0$ & $1$ & $2$ & $3$ & $4$ & $5$ & $6$ & $7$ & $8$ & $9$  \\
 \hline
 $0$ & $\frac{1}{4}$ & $\frac{1}{4}$ & $\frac{3}{8}$ & $ \frac{15}{32}$ & $\frac{35}{64}$ & $\frac{315}{512}$ & $  \frac{693}{1024}$ & $\frac{3003}{4096}$ & $  \frac{6435}{8192}$ & $\frac{109395}{131072}$ \\
 \hline
 $1$ & $\frac{1}{4}$ & $\frac{1}{4}$ & $\frac{1}{8}$ & $  \frac{3}{32}$ & $\frac{5}{64}$ & $\frac{35}{512}$ & $  \frac{63}{1024}$ & $\frac{231}{4096}$ & $  \frac{429}{8192}$ & $\frac{6435}{131072}$ \\
 \hline
 $2$ & $\frac{3}{8}$ & $\frac{1}{8}$ & $\frac{1}{16}$ & $  \frac{3}{64}$ & $\frac{5}{128}$ & $\frac{35}{1024}$ & $  \frac{63}{2048}$ & $\frac{231}{8192}$ & $  \frac{429}{16384}$ & $\frac{6435}{262144}$ \\
 \hline
 $3$ & $\frac{15}{32}$ & $\frac{3}{32}$ & $\frac{3}{64}$ & $  \frac{9}{256}$ & $\frac{15}{512}$ & $\frac{105}{4096}$ & $  \frac{189}{8192}$ & $\frac{693}{32768}$ & $  \frac{1287}{65536}$ & $ \frac{19305}{1048576}$ \\
 \hline
 $4$ & $\frac{35}{64}$ & $\frac{5}{64}$ & $\frac{5}{128}$ & $  \frac{15}{512}$ & $\frac{25}{1024}$ & $\frac{175}{8192}  $ & $\frac{315}{16384}$ & $\frac{1155}{65536}$ & $  \frac{2145}{131072}$ & $ \frac{32175}{2097152}$ \\
 \hline
 $5$ & $\frac{315}{512}$ & $\frac{35}{512}$ & $  \frac{35}{1024}$ & $\frac{105}{4096}$ & $  \frac{175}{8192}$ & $\frac{1225}{65536}$ & $  \frac{2205}{131072}$ & $\frac{8085}{524288}$ & $  \frac{15015}{1048576}$ & $\frac{225225}{16777216}$ \\
 \hline
 $6$ & $\frac{693}{1024}$ & $\frac{63}{1024}$ & $  \frac{63}{2048}$ & $\frac{189}{8192}$ & $  \frac{315}{16384}$ & $\frac{2205}{131072}$ & $  \frac{3969}{262144}$ & $\frac{14553}{1048576}$ & $  \frac{27027}{2097152}$ & $\frac{405405}{33554432}$ \\
 \hline
 $7$ & $\frac{3003}{4096}$ & $\frac{231}{4096}$ & $  \frac{231}{8192}$ & $\frac{693}{32768}$ & $  \frac{1155}{65536}$ & $\frac{8085}{524288}$ & $  \frac{14553}{1048576}$ & $\frac{53361}{4194304}$ & $  \frac{99099}{8388608}$ & $\frac{1486485}{134217728}  $ \\
 \hline
 $8$ & $\frac{6435}{8192}$ & $\frac{429}{8192}$ &
   $\frac{429}{16384}$ & $\frac{1287}{65536}$ &
   $\frac{2145}{131072}$ & $\frac{15015}{1048576}$ &
   $\frac{27027}{2097152}$ & $\frac{99099}{8388608}$ &
   $\frac{184041}{16777216}$ & $\frac{2760615}{268435456}$ \\
 \hline
 $9$ & $\frac{109395}{131072}$ & $\frac{6435}{131072}$ & $  \frac{6435}{262144}$ & $\frac{19305}{1048576}$ & $  \frac{32175}{2097152}$ & $\frac{225225}{16777216}$ & $  \frac{405405}{33554432}$ & $ \frac{1486485}{134217728}  $ & $\frac{2760615}{268435456}$ & $  \frac{41409225}{4294967296}$ \\
 \hline
\end{tabular}
  \caption{Coefficients $F_{nm}$ for computation of the interaction energy, see Eq.~\eqref{Enm}, \eqref{Enm_dim}.}\label{G1G2}
\end{table}

\begin{table}[!hbt]
  \centering
 \begin{tabular}{|c|c|c|c|c|c|c|c|c|c|c|}
 \hline
 $m \backslash n$
& $ 0$ & $1$ & $2$ & $3$ & $4$ & $5$ & $6$ & $7$ & $8$ & $9$ \\
 \hline
 $0$ & $-\frac{1}{4}$ & $-\frac{1}{4}$ & $-\frac{3}{8}$ & $  -\frac{15}{32}$ & $-\frac{35}{64}$ & $-\frac{315}{512}  $ & $-\frac{693}{1024}$ & $-\frac{3003}{4096}$ & $  -\frac{6435}{8192}$ & $-\frac{109395}{131072}$ \\
 \hline
 $1$ & $-\frac{1}{4}$ & $\frac{1}{4}$ & $\frac{3}{8}$ & $  \frac{15}{32}$ & $\frac{35}{64}$ & $\frac{315}{512}$ & $  \frac{693}{1024}$ & $\frac{3003}{4096}$ & $  \frac{6435}{8192}$ & $\frac{109395}{131072}$ \\
 \hline
 $2$ & $-\frac{3}{8}$ & $\frac{3}{8}$ & $\frac{15}{16}$ & $  \frac{105}{64}$ & $\frac{315}{128}$ & $  \frac{3465}{1024}$ & $\frac{9009}{2048}$ & $  \frac{45045}{8192}$ & $\frac{109395}{16384}$ & $  \frac{2078505}{262144}$ \\
 \hline
 $3$ & $-\frac{15}{32}$ & $\frac{15}{32}$ & $\frac{105}{64}$ & $  \frac{945}{256}$ & $\frac{3465}{512}$ & $  \frac{45045}{4096}$ & $\frac{135135}{8192}$ & $  \frac{765765}{32768}$ & $\frac{2078505}{65536}$ & $  \frac{43648605}{1048576}$ \\
 \hline
 $4$ & $-\frac{35}{64}$ & $\frac{35}{64}$ & $\frac{315}{128}  $ & $\frac{3465}{512}$ & $\frac{15015}{1024}$ & $  \frac{225225}{8192}$ & $\frac{765765}{16384}$ & $  \frac{4849845}{65536}$ & $ \frac{14549535}{131072}$ & $  \frac{334639305}{2097152}$ \\
 \hline
 $5$ & $-\frac{315}{512}$ & $\frac{315}{512}$ & $  \frac{3465}{1024}$ & $\frac{45045}{4096}$ & $  \frac{225225}{8192}$ & $\frac{3828825}{65536}$ & $  \frac{14549535}{131072}$ & $\frac{101846745}{524288}  $ & $\frac{334639305}{1048576}$ & $  \frac{8365982625}{16777216}$ \\
 \hline
 $6$ & $-\frac{693}{1024}$ & $\frac{693}{1024}$ & $  \frac{9009}{2048}$ & $\frac{135135}{8192}$ & $  \frac{765765}{16384}$ & $\frac{14549535}{131072}$ & $  \frac{61108047}{262144}$ & $ \frac{468495027}{1048576}  $ & $\frac{1673196525}{2097152}$ & $  \frac{45176306175}{33554432}$ \\
 \hline
 $7$ & $-\frac{3003}{4096}$ & $\frac{3003}{4096}$ & $  \frac{45045}{8192}$ & $\frac{765765}{32768}$ & $  \frac{4849845}{65536}$ & $\frac{101846745}{524288}$ & $  \frac{468495027}{1048576}$ & $  \frac{3904125225}{4194304}$ & $  \frac{15058768725}{8388608}$ & $  \frac{436704293025}{134217728}$ \\
 \hline
 $8$ & $-\frac{6435}{8192}$ & $\frac{3003}{4096}$ &
   $\frac{45045}{8192}$ & $\frac{765765}{32768}$ &
   $\frac{4849845}{65536}$ & $\frac{101846745}{524288}$ &
   $\frac{468495027}{1048576}$ &
   $\frac{3904125225}{4194304}$ &
   $\frac{15058768725}{8388608}$ &
   $\frac{436704293025}{134217728}$ \\
 \hline
 $9$ & $-\frac{109395}{131072}$ & $\frac{109395}{131072}$ & $  \frac{2078505}{262144}$ & $\frac{43648605}{1048576}$ & $  \frac{334639305}{2097152}$ & $  \frac{8365982625}{16777216}$ & $  \frac{45176306175}{33554432}$ & $  \frac{436704293025}{134217728}$ & $  \frac{1933976154825}{268435456}$ & $  \frac{63821213109225}{4294967296}$ \\
 \hline
\end{tabular}
\caption{Coefficients $H_{nm}$ for computation of the interaction energy, see Eq.~\eqref{Enm}, \eqref{Enm_dim}.}\label{CnCm}
\end{table}
\end{widetext}

In the special case of $m=0$ we get the following expression for the dimensionless interaction energy as a function of $\tilde{d}$ and $\gamma_1$:
\begin{equation}\label{Enm_sc1}
\Delta \tilde{E}_{n0} = -\frac{a_1 a_2\pi}{\tilde{d}^{|n-1|+2}}(1-\nu)L_{n}\cos(n\gamma_1) ,
\end{equation}
with a single coefficient $L_n$. Similarly, when $n=0$ we get:
\begin{equation}\label{Enm_sc2}
\Delta \tilde{E}_{0m} = -\frac{a_1 a_2\pi}{\tilde{d}^{|m-1|+2}}(1-\nu)L_{m}\cos(m\gamma_2).
\end{equation}
In order to incorporate these results in Tables~\ref{G1G2}-\ref{CnCm} we set $F_{n0}=-H_{n0}=L_n/2$ and $F_{0m}=-H_{0m}=L_m/2$, and this takes care of the coefficient $(1-\nu)$ that includes Poisson's ratio.

\begin{figure}[t]
\includegraphics[width=\columnwidth]{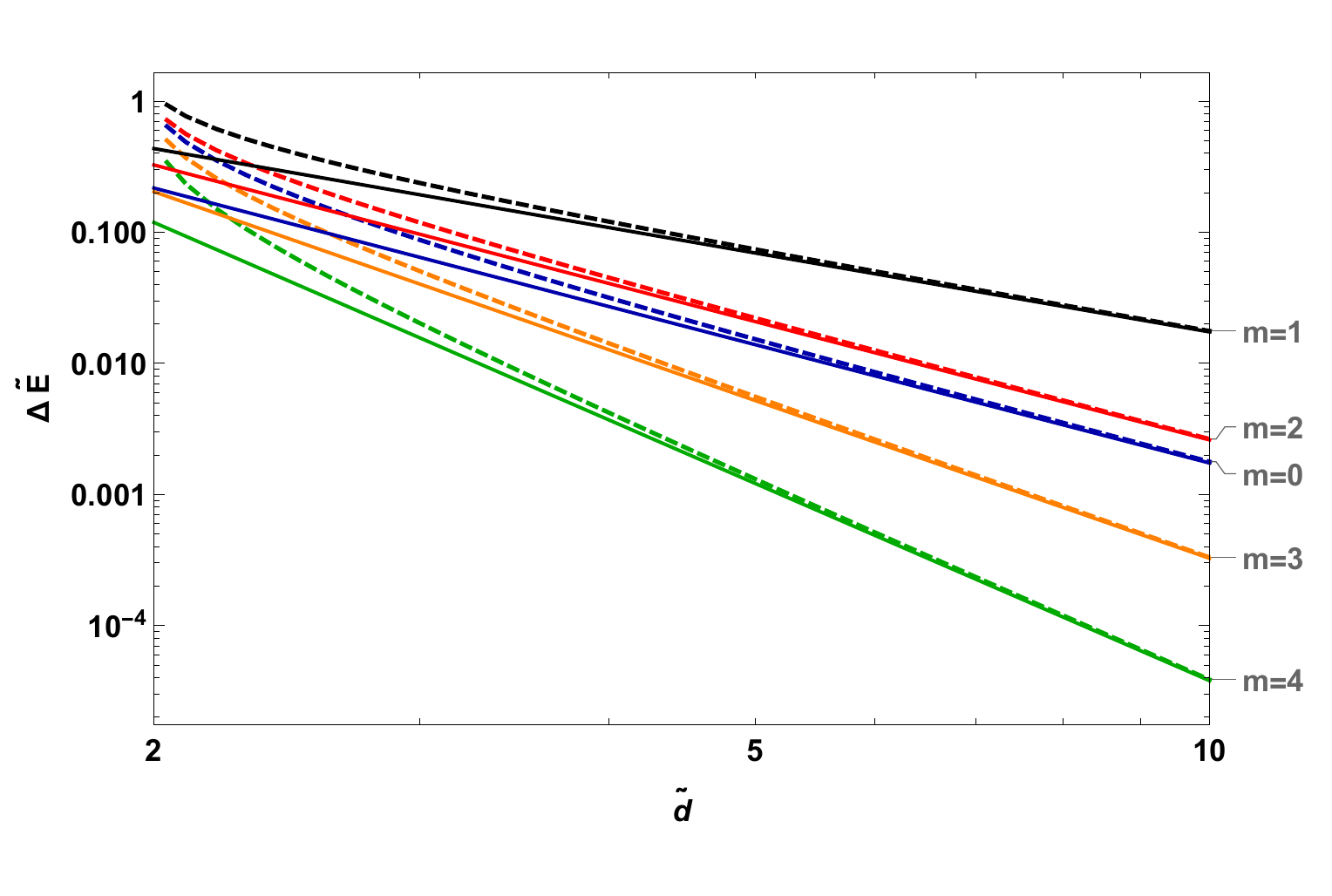}
\caption{Non-dimensional interaction energy $\Delta \tilde{E}$ between two active discs applying radial forces $\tilde{f}_1\equiv 1$ and $\tilde{f}_2=\cos (m \theta_2)$ with $0\leq m\leq4$ vs. the non-dimensional distance $\tilde{d}$ between their centers. Dashed lines represent the exact numerical solution and solid lines represent the analytical approximation of Eq.~\eqref{Enm}. The value of Poisson's ratio was set to $\nu=0.45$.}
\label{Interaction energy vs distance}
\end{figure}

Figure \ref{Interaction energy vs distance} shows the normalized interaction energy $\Delta \tilde{E}$ computed numerically (exact) and analytically (approximate) vs. the normalized distance $\tilde{d}$ between two active discs applying forces  $\tilde{f}_1(\theta_1)=1$ and $\tilde{f}_2(\theta_2)=cos (m\theta_2)$ for $0 \leq m \leq 4$. The analytical expressions were computed using Eq.~\eqref{Enm} and Tables \ref{G1G2}-\ref{CnCm}. As seen in the figure, for large $\tilde{d}$ the approximation of the interaction energy to the leading term is enough to capture the rate of the decay of interaction energy with $\tilde{d}$. As seen in Eq.~\eqref{Enm} the interaction energy decays with distance as $\tilde{d}^{-q}$ where $q=|m-1|+|n-1|+1$ is an integer number, which grows with the order of the term for $n,m>0$. When the distance between the active discs is small, a good approximation requires additional terms, and thus the numerical solution may be more practical.

\section{Discussion}\label{discussion}

\subsection{Dependence on phase angles}

\begin{figure}[h]
\includegraphics[width=0.8\columnwidth]{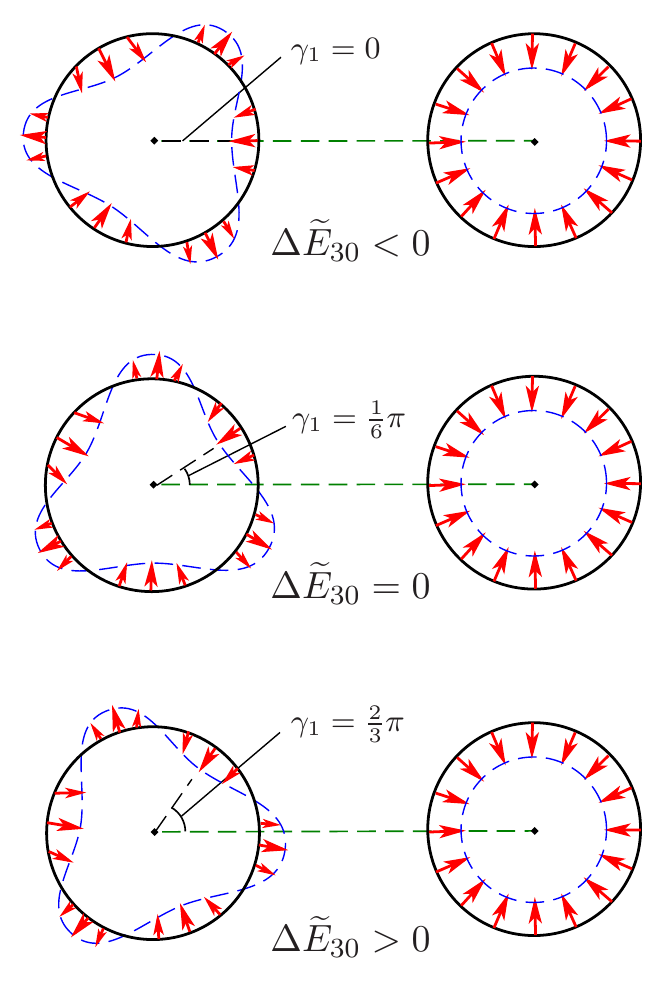}
\caption{System of two active discs applying forces that result in the interaction energy given by Eq.~\eqref{Enm_sc1}, demonstrating the change of sign of $\Delta \tilde{E}_{30}$ as the phase angle $\gamma_1$ changes.}
\label{Ie 02 03}
\end{figure}

In Fig.~\ref{Ie 02 03} we demonstrate the dependence of the interaction energy on the phase angle. Specifically, we show a pair of discs with an isotropic contractile force ($m=0$) on disc 2, and a simple anisotropic forces ($n=3$) on disc 1. We show different phase angles $\gamma_1$ of the force on disc 1 that give repulsive (positive), attractive (negative), and zero interaction between the discs. 

\begin{figure}[!htb]
\centering
\includegraphics[width=\columnwidth]{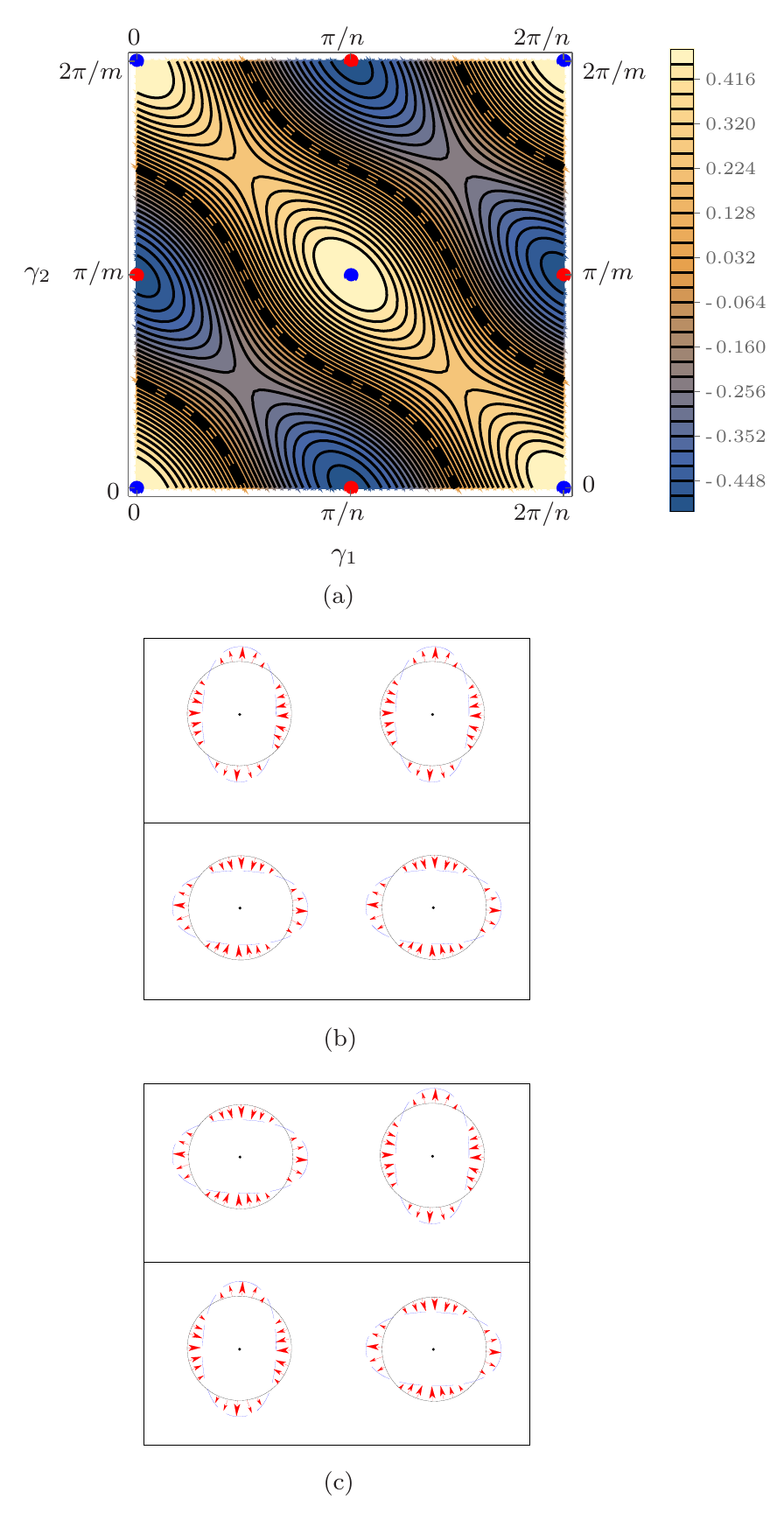}
 \caption{(a) Typical plot of the interaction energy $\Delta \tilde{E}_{nm}$ for $n,m>0$. Here $n=m=1$ and $\nu=0.45$. Thick black lines correspond to $\Delta \tilde{E}_{nm}=0$, blue dots correspond to local maxima ($\Delta \tilde{E}_{nm}>0$, i.e. repulsion), red dots to local minima ($\Delta \tilde{E}_{nm}<0$, i.e. attraction). (b-c) Mutual orientations of active discs with $n=m=2$ at the local maxima (b), and minima (c) of the interaction energy $\Delta \tilde{E}_{22}$.}
\label{g1g2tot}
\end{figure}

As one may expect, the interaction energies are generally periodic in the phases $\gamma_1$, $\gamma_2$ since the forces are periodic; phase angles $\gamma_1=\frac{2 \pi}{n}$ and $\gamma_2=\frac{2 \pi}{m}$ are equivalent to $\gamma_1=\gamma_2=0$. Indeed, all terms in Eq.~\eqref{Enm} include the functions $\cos(n \gamma_1 \pm m \gamma_2)$. Since both functions and their derivatives are continuous and since for any $n,m>0$, the coefficients $F_{nm}$, $H_{nm}$ are positive, extremum points for any $\Delta E_{nm}$ will appear at $\left\{\gamma_1=k_1\frac{\pi}{n},\gamma_2=k_2\frac{\pi}{m}\right\}$ with integer $k_1$ and $k_2$, where local maxima correspond to $k_1+k_2=2s$ and local minima to $k_1+k_2=2s+1$ with $s$ integer, see Figure~\ref{g1g2tot}. It may also be seen that for some values of $\gamma_1$ and $\gamma_2$ the interaction energy is positive while for other values it will become negative. In other words, in some orientations the active discs will attract while in other orientations they will repel. The exact position of the $\Delta E_{nm}=0$ lines in the $(\gamma_1,\gamma_2)$ plane depends on the expressions for $F_{nm}$ and $H_{nm}$ and thus will be different for every $n$ and $m$. In addition, as seen from Eq.~\eqref{Enm} the value of Poisson's ratio $\nu$ also affects the position of the $\Delta E_{nm}=0$ lines in the $(\gamma_1,\gamma_2)$ plane.

\subsection{Comparison to linear force dipoles}

We mentioned earlier that we focus here on linear (Hookean) elasticity and small deformations. This allows us to use the superposition method for the force and displacements fields. Any periodic force distribution $f_i(\theta_i)$  may be written as a multipole expansion using the Fourier series:
\begin{equation}\label{Fourier}
f_i(\theta_i)=\sum_{n} \left\{ C_{n,i} \cos\left[n \left(\theta_i-\gamma_{n,i}\right)\right] \right\}.
\end{equation}

\begin{figure}[t]
\includegraphics[width=0.45\textwidth]{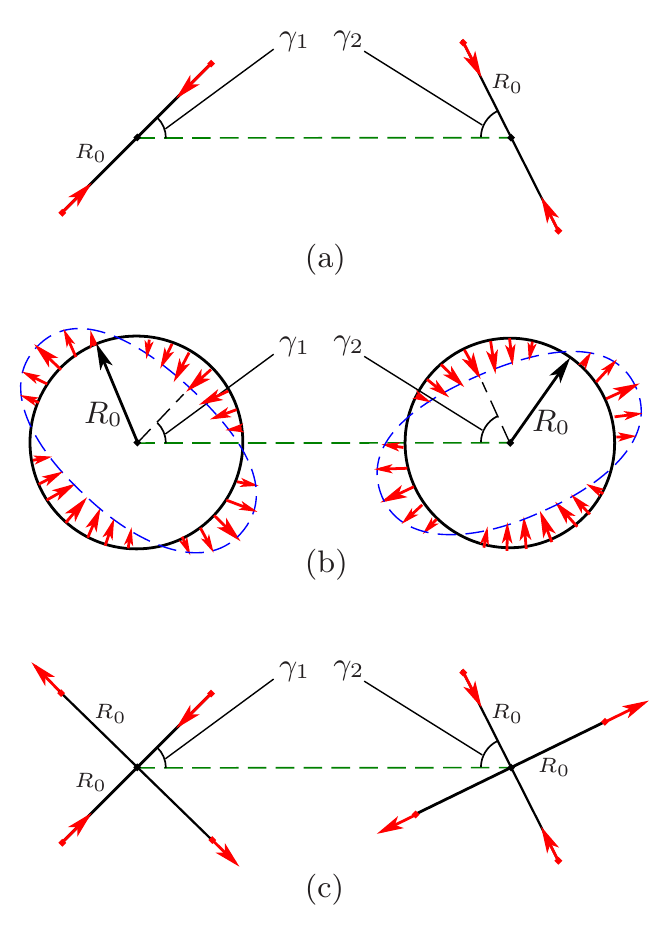}
\caption{Analogy of active discs to linear force dipoles. (a) Pair of linear force dipoles. (b) Pair of active discs applying mode $n=2$ with phase angles $\gamma_1$ and $\gamma_2$. (c) Analogous system, in which each disc is replaced by two orthogonal linear force dipoles.}
\label{dd}
\end{figure}

Mode $n=2$ in this series is related to the linear force dipole, since it has an axis along which the forces concentrate. The interaction energy between two linear force dipoles separated by a distance $d$ and oriented at angles $\gamma_1$ and $\gamma_2$ with respect to the axis between them (see Fig.~\ref{dd}a) is given by~\cite{Bischofs2006}
\begin{align}\label{aldie2}
  \Delta E &= - \frac{F^2 R_0^2}{4 \pi G d^3} \\
   &\cdot \left\{ 2 (1-\nu)\left[1 + 3( \cos{2\gamma_1}+\cos{2\gamma_2})\right] \right. \nonumber\\
  &\left. +15 \nu \cos{(2(\gamma_1-\gamma_2))} +(2-\nu) \cos{(2(\gamma_1+\gamma_2))} \right\} \nonumber ,
\end{align}
where each linear force dipole consists of two point forces $F$ separated by a distance $2R_0$. The interaction energy we obtain from Eq.~\eqref{Enm_dim} for two discs applying only the $m=n=2$ mode is:
\begin{align}\label{2modeie}
  \Delta E&= -\frac{\pi R_0^4 A_1 A_2}{16 G d^3} \\
  &\cdot \left\{ (2-\nu) \cos{[2(\gamma_1+\gamma_2)]} + 15 \nu \cos{[2(\gamma_1-\gamma_2)]} \right\} . \nonumber
\end{align}

These two expressions~(\ref{aldie2},\ref{2modeie}) have the same $1/d^3$ dependnce on distance, but a different dependence on the relative orientations $\gamma_1$,$\gamma_2$. To make the connection between the two cases, we note that mode $n=2$ of an active disc includes not only a concentration of two opposing contractile (inward) forces, but also a pair of extensile (outward) forces, see Fig.~\ref{dd}b. Thus we compare each $n=2$ active disc not to a contractile linear force dipole, but to a combination of two orthogonal linear force dipoles, one applying contraction forces and another applying expansion forces, see Fig.~\ref{dd}c. We evaluated the interaction energy between two such double force dipoles using Eq.~\eqref{aldie2}. We found that the angular dependence exactly matches that in the interaction between two $n=2$ active discs as given by Eq.~\eqref{2modeie}. The prefactors are matched if we set the magnitude of the force distributions to $A_1=A_2=\frac{4 F}{\pi R_0}$.

\begin{figure}[t]
\includegraphics[width=0.45\textwidth]{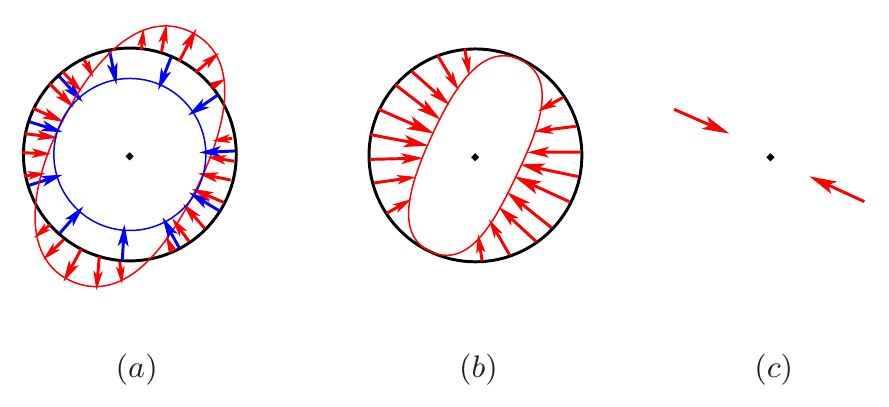}
\caption{Approximation of linear force dipole by superposition of $n=0$ and $n=2$ modes. (a) $n=0$ (blue) and $n=2$ (red) modes in the same coordinate system. (b) Result of the superposition of both modes. (c) The analogous linear force dipole.}
\label{lda}
\end{figure}

We can also perform a comparison between our active discs and the previously-studied linear force dipoles from the opposite perspective. Namely, to consider the combination of Fourier modes on a disc that best describes a linear force dipole, and compare the interaction energies obtained from each of the two models. We describe a linear force dipole by adding the isotropic mode $n=0$ to the mode $n=2$ on the disc in order to cancel the outward part of the radial force in the $n=2$ mode. Namely we take $f_i(\theta_i) = A_0 + A_2 \cos[2(\theta_i-\gamma_i)]\}$, see Fig.~\ref{lda}. We note that in this case the spatial decay of the contributions to the interaction energy is identical for all four interactions terms (0,0), (0,2), (2,0), and (2,2), since $\Delta E \propto d^{-q}$ with $q=|n-1|+|m-1|+1$. Using our far-field approximation Eq.~\eqref{Enm_dim}, we get the following full dependence on the phase angles $\gamma_1$ and $\gamma_2$:
\begin{align}\label{aldie1}
  \Delta E&=\Delta E_{00} + \Delta E_{20} + \Delta E_{02} + \Delta E_{22} \\ &= - \frac{\pi R_0^4}{16 G d^3} \left\{ 8 (1-\nu) A_0^2 \right. \nonumber\\  &\left. + 12(1-\nu) A_0 A_2 \left[\cos{(2\gamma_1)}+\cos{(2\gamma_2)}\right] \right. \nonumber\\  &\left. + A_2^2 [15 \nu \cos{(2(\gamma_1-\gamma_2))} + (2-\nu) \cos{(2(\gamma_1+\gamma_2))} ] \right\}. \nonumber
\end{align}
We obtain a very similar angular dependence as for the interaction between linear force dipoles. However, for this to perfectly match the expression for the interaction between two linear force dipoles~\eqref{aldie2}, we would need to set $A_2=2A_0= \frac{2F}{\pi R_0}$, while the geometry presented in Fig.~\ref{lda} implies that $A_2=A_0$ so that the force will vanish in the transverse direction. Nonetheless, we have established a strong connection between the previously-studied model of linear force dipoles and our model of anisotropically-contracting active discs.

Modes $n>2$ can be thought of as higher-order multipoles. Mode $n=0$ is also related to a force dipole since it is an isotropic collection of pairs of point forces around the disc. In contrast to all other modes with zero net force, mode $n=1$ may be thought of as a force monopole; the net force in this case does not vanish, and it is thus less relevant to non-motile contractile cells. This is similar but not fully equivalent to electric multipoles, since in electrostatics the charges are scalar while in elasticity the forces are vectorial and thus even if we restrict ourselves to radial forces, we need to sum them vectorially in order to understand the long-range effect of these forces. Thus n=0 has a meaning for forces but not for electric charges. See ~\cite{Shokef2012, shokeferr, whycellscare, 1367-2630-19-6-063011}

Due to the superposition principle the interaction energy of the system of two active disc may be written as:
\begin{equation}\label{total interaction energy series}
  \Delta \tilde{E} = \sum_{n \ne m} C^{(1)}_{n} C^{(2)}_{m} \Delta \tilde{E}_{nm,tot} ,
\end{equation}
where the coefficients $C^{(1)}_{n}$ and $C^{(2)}_{m}$ are the amplitudes of different modes of the radial forces created by the active discs $1$ and $2$ respectively. From Eq.~\eqref{total interaction energy series} it follows that knowledge of the interaction energies of different modes $\Delta \tilde{E}_{mn,tot}$ makes it possible to evaluate the interaction energies of arbitrary force distributions $\tilde{f}_1$ and $\tilde{f}_2$.

\section{Conclusions}\label{conclusions}

We modeled live cells as discs resting on the surface of a semi-infinite substrate with linear (Hookean) properties, and applying on the substrate anisotropic forces directed to their centers along their edges. We found the interaction energy between every two cells to be inversely proportional to integer powers of the distance between them. This power depends on the Fourier modes, $n$ and $m$ of the forces applied by the active discs. We also found that the interaction energy is proportional to a linear combination of the functions $\cos(n\gamma_1 +m\gamma_2)$ and $\cos(n\gamma_1 -m\gamma_2)$, where $\gamma_1$ and $\gamma_2$ are the phase angles of the active forces applied by each active disc. The linearity of the equations makes it possible to evaluate the interaction energy for more complex force distributions by combining the results that we presented. We deliberately simplified the biological setup to the tractable geometry of circular discs. In principle, our approach could be extended to describe cells with arbitrary shape, and not only the arbitrary azimuthal force distribution that we studied here.

Biological cells have finite thickness and finite stiffness, thus the application of forces to the substrate creates deformation fields in the cells as well and not only in the substrate, and it would be interesting to take into account the additional elastic energy stored in the cells in order to evaluate the total elastic energy of the system. Following our work on zero-thickness active discs on a semi-infinite elastic substrate, it would be interesting to consider finite-thickness active discs with elastic properties that differ from those of the substrate. Taking into account the difference in elastic properties between the substrate and the cells will help understand the behavior of cells plated on gels with different rigidities, as tested experimentally. Finally, taking into account the nonlinearity of the substrate~\cite{Winer2009, lesman_arxiv, Shokef2012, shokeferr}, our analytical procedure could clearly not be employed, and this could be an interesting direction for future numerical research.

\begin{acknowledgments}

We thank Aparna Baskaran, Haim Diamant, Ben Hancock, Ayelet Lesman, Sam Safran and Chaviva Sirote for helpful discussions. This work was partially supported by the Israel Science Foundation grant No. 968/16, by a grant from the United States-Israel Binational Science Foundation and by a grant from the Ela Kodesz Institute for Medical Physics and Engineering.

\end{acknowledgments}

\end{document}